\newif\ifConf
\Conffalse

\ifConf
\documentclass[envcountsect]{llncs}

\usepackage{amsmath}
\usepackage{amssymb}
\usepackage{times}
\usepackage{diagrams}

\newcommand{\citeyear}[1]{\cite{#1}}

\else

\documentclass[11pt]{article}

\usepackage{chicagor}
\usepackage{amsmath}
\usepackage{amssymb}
\usepackage{times}
\usepackage{diagrams}
\usepackage{theorems}

\fi

\allowdisplaybreaks[2]

\newcommand{\thmcolon}{}

\ifConf\else

\newtheorem{THEOREM}{Theorem}[section]
\newtheorem{LEMMA}[THEOREM]{Lemma}
\newtheorem{COROLLARY}[THEOREM]{Corollary}
\newtheorem{PROPOSITION}[THEOREM]{Proposition}
\newtheorem{DEFINITION}[THEOREM]{Definition}
\newtheorem{CLAIM}[THEOREM]{Claim}
\newtheorem{EXAMPLE}[THEOREM]{Example}
\newtheorem{REMARK}[THEOREM]{Remark}

\newenvironment{theorem}{\begin{THEOREM} \thmcolon  }%
                        {\end{THEOREM}}
\newenvironment{lemma}{\begin{LEMMA} \thmcolon  }%
                      {\end{LEMMA}}
                          {\end{COROLLARY}}
                            {\end{PROPOSITION}}
                            { \end{DEFINITION}}
                            {\end{CLAIM}}

\newenvironment{example}{\begin{EXAMPLE} \thmcolon  \rm}%
                            { \wbox\end{EXAMPLE}}
\newenvironment{example*}{\begin{EXAMPLE} \thmcolon  \rm}%
                            {\end{EXAMPLE}}
                            { \wbox\end{REMARK}}
\newenvironment{remark*}{\begin{REMARK} \thmcolon  \rm}%
                            {\end{REMARK}}

\def\squareforqed{\hbox{\rlap{$\sqcap$}$\sqcup$}}
\def\wbox{\ifmmode\squareforqed\else{\unskip\nobreak\hfil
\penalty50\hskip1em\null\nobreak\hfil\squareforqed
\parfillskip=0pt\finalhyphendemerits=0\endgraf}\fi}
\newenvironment{proof}{\noindent{\it Proof.}}
                      {\wbox\vspace{0.1in}}

\fi

\renewcommand{\phi}{\varphi}
\newcommand{\<}{\langle}
\renewcommand{\>}{\rangle}
\renewcommand{\emptyset}{\varnothing}

\newcommand{\summ}{\downarrow}
\newcommand{\Hom}{\mathrm{Hom}}

\newcommand{\cK}{\mathcal{K}}
\newcommand{\cS}{\mathcal{S}}
\newcommand{\cM}{\mathcal{M}} 
\newcommand{\cI}{\mathcal{I}}
\newcommand{\cJ}{\mathcal{J}}

\newcommand{\cF}{\mathcal{F}}
\newcommand{\bbN}{\mathbb{N}}
\newcommand{\cat}[1]{\mathbf{#1}}

\newcommand{\fS}{S}     %
\newcommand{\fK}{K}     %
\newcommand{\fP}{P}     %
\newcommand{\fT}{T}     %
\newcommand{\fU}{U}     %
\newcommand{\fC}{C}     %
\newcommand{\fPp}{P'}     %
\newcommand{\fTp}{T'}     %
\newcommand{\fUp}{U'}     %
\newcommand{\fCp}{C'}     %

\newcommand{\ostar}
  {\ensuremath{\settowidth{\dimen7}{\mbox{$\odot$}}\makebox[0pt][l]{$\odot$}\makebox[\dimen7]{\mbox{$\ast$}}}}

\newcommand{\sini}{{\scriptscriptstyle\rm ini}}
\newcommand{\sfin}{{\scriptscriptstyle\rm fin}}

\ifConf

 \title{On Partially Additive Kleene Algebras}
\author{Riccardo Pucella}
\institute{Cornell University\\
Ithaca, NY 14853 USA\\
riccardo@cs.cornell.edu}

\else

 \title{On Partially Additive Kleene Algebras\thanks{A preliminary
 version of this paper appears in the proceedings of the 8th
 International Conference on Relational Methods in Computer Science
 (RelMiCS 8).}}
\author{Riccardo Pucella\\
Cornell University\\
Ithaca, NY 14853 USA\\
riccardo@cs.cornell.edu}
\date{}

\fi

\begin{document}

\maketitle

\begin{abstract}
We define the notion of a partially additive Kleene algebra, which is
a Kleene algebra where the $+$ operation need only be partially
defined. These structures formalize a number of examples that cannot
be handled directly by Kleene algebras. We relate partially additive
Kleene algebras to existing algebraic structures, by exhibiting
categorical connections with Kleene algebras, partially additive
categories,  and closed semirings.
\end{abstract}

\section{Introduction}

Kleene algebras are algebraic structures that capture a natural form
of iteration. Formally, a Kleene algebra is an idempotent semiring
with a unary $*$ operation; it is this operation that is axiomatized
using properties one expects from iteration. Kleene algebras are
well-understood, having been studied since at least the work of Conway
\citeyear{r:conway71}, and provide a foundation for many program
logics. The goal of this paper is to extend the theory of Kleene
algebras to capture situations where it is possible to define a
reasonable definition of iteration without there being a complete
Kleene algebra structure. One example, studied by Kozen
\citeyear{r:kozen98}, is the set of nonsquare matrices over a Kleene
algebra, which does not have a multiplication operation defined for
all pairs of matrices. This example can be studied by a form of Kleene
algebra with a partially defined multiplication operation.  
We are interested in situations that require a different form of
partiality, namely, where the addition operation is partial. An
example where a partially defined addition operation is needed is
given (essentially) by the set of partial functions on a set $S$, that
is, the partial functions $S\rightarrow S$. We can define
multiplication of two partial functions by composition. Similarly, we
can define the sum of two partial functions $f$ and $g$ as $f\cup g$
(when considered as relations between arguments and results) provided
that the values of $f$ and $g$ agree on their domain of definition;
thus, addition of partial functions is only partially
defined. Finally, we can define $f^*$ as a form of reflexive
transitive closure of $f$.
Of course, partial functions can be studied using a number of
approaches. This simply shows that we can bring them (and other
examples) under the umbrella of a well-studied algebraic theory such
as that of Kleene algebras.

Addition and multiplication having different properties in Kleene
algebras, a theory for a partially defined addition operation is
different than the theory for a partially defined multiplication
operation. This means that the development of Kozen
\citeyear{r:kozen98} for a partially defined multiplication is not
directly applicable. Of course, the need for a partially defined
addition has long been recognized, and a common way to deal with this
issue is to consider a Kleene algebra with a special element $\star$
to represent the value ``undefined''. When the sum of two elements is
meant to be undefined, it is assigned the value $\star$, and applying
any operation to a $\star$ argument results in $\star$. This is
essentially the approach used by Kozen \citeyear{r:kozen03a} in a
recent series of papers to perform dataflow analyses using Kleene
algebras. While this method has the definite advantage of being simple
and does not require any extension to the theory of Kleene algebras,
it may not be always adequate. Among other things, it is not a
\emph{universal} way to add an undefined value to a Kleene
algebra. For instance, we could imagine adding a number of undefined
values, one per pair of elements whose sum is undefined, and defining
addition appropriately for those values. There are properties of
Kleene algebras extended with many undefined values that do not hold
when a single undefined value is used, especially as it pertains to
homomorphisms between Kleene algebras.  What we want, for the purpose
of developing a theory of Kleene algebras with a partially defined
addition, is to identify the ``minimal'' way in which undefined values
can be added, for an appropriate sense of minimality. It is not clear,
a priori, that there is such a minimal extension.

To establish that such a minimal extension indeed exists, we proceed
as follows. We define a primitive notion of \emph{partially additive}
Kleene algebra, which is a Kleene algebra where the $+$ operation is
partial, rather than total. We establish the basic result that every
star-continuous partially additive Kleene algebra (a class of algebras
where the $*$ operation satisfies some additional and natural
properties) can be embedded in a total Kleene algebra, and that this
embedding is universal. Thus, there is a universal way to complete
(star-continuous) partially additive Kleene algebras. This universal
completion yields the minimal extension alluded to above.
The remainder of the paper describes the relationship between
partially additive Kleene algebras and other algebraic structures,
with an eye towards showing that such Kleene algebras fit naturally
within existing structures. First, we relate partially additive Kleene
algebras with another structure that has been used to reason
algebraically about iteration in the context of categorical semantics
of programming languages, namely partially additive idempotent
semirings and their generalization as partially additive categories
\cite{r:manes86}. We exhibit an adjunction between the category of
star-continuous partially additive Kleene algebras and the category of
partially additive idempotent semirings. Furthermore, we show that
star-continuous partially additive Kleene algebras can be completed
into closed semirings \cite{r:aho75} in two distinct ways that
nevertheless yield isomorphic closed semirings. This is done by
exhibit two adjunctions between the category of star-continuous
partially additive Kleene algebras and the category of closed
semirings.

\ifConf
\else
We assume a working knowledge of category theory, at the level of the
first five chapters of MacLane \citeyear{r:maclane71}. 
\fi

\section{Partially Additive Kleene Algebras}\label{s:pka}

We take as a starting point the definition of a Kleene algebra due to
Kozen \citeyear{r:kozen94}, who gives a finitary axiomatization using
implicational equations. 

A \emph{partially additive Kleene algebra} is a set $\cK$ equipped
with a partial binary operation $+$, a total binary operation $\cdot$, 
a total unary operation $^*$, and constants $0$ and $1$, satisfying a 
number of axioms. We say the pair $(a,b)$ is \emph{summable}
(written $a\summ b$) if $a+b$ is defined. The axioms are as follows:
\begin{gather}
\text{if $x\summ y$, $(x+y)\summ z$, and $y\summ z$ then  $x\summ(y+z)$ and 
$(x+y)+z = x+(y+z)$}\\
\text{if $x\summ y$ then $y\summ x$ and $x+y=y+z$}\\
\text{$x\summ 0$ and $x+0=x$}\\
\text{$x\summ x$ and $x+x=x$}\\
x\cdot(y\cdot z) = (x\cdot y)\cdot z\\
1\cdot x=x\\
x\cdot 1=x\\
\text{if $x\summ y$ then $(z\cdot x)\summ(z\cdot y)$ and $z\cdot(x+y)=z\cdot x+z\cdot y$}\\
\text{if $x\summ y$ then $(x\cdot z)\summ(y\cdot z)$ and $(x+y)\cdot z=x\cdot z+y\cdot z$}\\
0\cdot x=0\\
x\cdot 0=0\\
\label{e:xstar}
1\le x^*\\
x\cdot x^*\le x^*\\
x^*\cdot x \le x^*\\
\text{if $a\cdot x\le x$ then $a^*\cdot x\le x$}\\
\text{if $x\cdot a\le x$ then $x\cdot a^*\le x$}
\end{gather}
As usual, every variable appearing in an axiom is implicitly
understood as universally quantified. The relation $\le$ in axioms
(12)--(16) is the natural partial order on $\cK$ induced by $+$: $a\le
b$ if and only if $a\summ b$ and $a+b=b$. It is straightforward to
check that $\le$ is symmetric (since $a\summ 0$), transitive (by the
associativity rules for summability), and antisymmetric (using the
idempotence of $+$). It is easy to check that $+$ captures a form of
least upperbound: if $a\le c$, $b\le c$, and $a\summ b$, then $a+b \le
c$ (using associativity rules for summability). We often drop the
$\cdot$ symbol for multiplication, simply writing $ab$ for $a\cdot
b$. For $b$ an element of a partially additive Kleene algebra, define
$b^0$ to be $1$, and $b^{n+1}$ to be $bb^{n}$.

Axioms (1)--(4) say that $(\cK,+,0)$ is an idempotent commutative
monoid, except that the $+$ operation is partial, while axioms
(5)--(9) say that $(\cK,\cdot,1)$ is a monoid, with the $\cdot$
operation distributing over $+$ whenever the required sums are
defined. With axioms (10)--(11), this makes $(\cK,+,\cdot,0,1)$ an
idempotent semiring with a partially defined $+$ operation. Axioms
(14)--(17) capture the properties of the Kleene star operation of
formal language theory, taking into account the partiality of $+$.
Most of the $\le$-related properties that hold in Kleene algebra also
hold here, even taking into account partial additivity (since $a\le b$
requires that $a\summ b$). For instance, we can check that for all
$n\ge 0$, $b^n\le b^*$ (using induction from
\eqref{e:xstar}). Similarly, $a^*a^*=a^*$, $(a^*)^*=a^*$, and if $a\le
b$, then $a^*\le b^*$. Rather than listing all the properties of
interest here, we refer to Kozen \citeyear{r:kozen94}.

Note that only $+$ is allowed to be partial in a partially
additive Kleene algebra. This lets us relate our results to partially
additive semirings in the next section, and also captures the kind of
examples we want to capture. Allowing $\cdot$ to be partial is
essentially an orthogonal feature. A form of Kleene algebra where the
$\cdot{}$ operation is partial is studied by Kozen \citeyear{r:kozen98}
under the name \emph{typed Kleene algebra}, where a type system is
used to control the partiality of multiplication. The definitions 
and results of this paper should carry over in a straightforward way 
to that setting. (We do not introduce a type system in this paper to 
control the partiality of addition, since it would make the framework 
too restrictive for the examples we want to capture.)

\begin{example}\label{x:total}
Every total Kleene algebra is a partially additive Kleene algebra
where the $+$ operation is in fact total. 
\end{example}

\begin{example}\label{x:monoid}
Any commutative monoid can be turned into a partially additive Kleene
algebra with a maximally undefined operation $+$. Let $(\cM,\cdot,1)$
be a commutative monoid. Let $\cM'=\cM\cup\{0,\star\}$, where $0$ and
$\star$ are new elements, and extend $\cdot$ to $\cM'$ by taking
$0\cdot a=a\cdot 0 = 0$ for all $a\in \cM'$, and $\star\cdot b=b\cdot
\star = \star$ for all $b\in
\cM'$, $b\ne 0$. Define $0^*=0$, and $b^*=\star$ (if $b\ne 0$). Let
$a+b$ be defined only if $a=b$, in which case $a+b=a$, if $b=0$, in
which case $a+b=a$, or if $b=\star$, in which case $a+b=b$. It is easy
to check that $(\cM',+,\cdot,*,0,1)$ forms a partially additive Kleene
algebra.
\end{example}

\begin{example}\label{x:pfn}
Let $\Sigma$ be a set of states. In much the same way that relations
are the basic example of Kleene algebras, the set of partial functions
from $\Sigma$ to $\Sigma$ is a basic example of a partially additive
Kleene algebra, where we interpret $f^*$ as repeatedly applying $f$
until the state is in a distinguished set
$\Omega\subseteq\Sigma$. Unfortunately, the example is not quite as
simple as in the total case. Intuitively, we need the partial
functions to record some information about how they have been
composed. Therefore, we look at strings over $\Sigma$, where we can
interpret a string $\sigma_1\dots \sigma_n$ as mapping $\sigma_1$ to
$\sigma_n$ (via $\sigma_2,\dots,\sigma_{n-1}$). Let $\epsilon$ be the
empty string. For a nonempty string $s=\sigma_1\dots\sigma_n$, let
$s_\sini$, the initial state of $s$, be $\sigma_1$, and $s_\sfin$, the
final state of $s$, be $\sigma_n$. Given two strings $s$ and $s'$, say
that $s$ is a generalized prefix of $s'$
if $s'=s_1t_1s_2t_2\dots s_n t_n$ and $s=s_1s_2\dots s_n$. The fusion
product $s\otimes s'$ of two strings is defined to be $\epsilon$ if
either $s$ or $s'$ is $\epsilon$, or if $s_\sfin\ne s'_\sini$. Otherwise, if
$s=\sigma_1\dots\sigma_m$ and $s'=\sigma'_1\dots\sigma'_n$, then
$s\otimes s'=\sigma_1\dots\sigma_m\sigma'_2\dots\sigma'_n$. Let
$\Omega\subseteq\Sigma$ be a fixed subset of $\Sigma$. A set
$A\subseteq \Sigma^*$ of strings is \emph{functional} if (1) for all
$s=\sigma_1\dots\sigma_n\in A$, if $\sigma_i\in\Omega$, then $i=n$
(only the last state of a string is in $\Omega$), (2) for all
$\sigma\in\Omega$, there is a string $s\in A$ with $s_\sini=\sigma$, and
(3) for all $\sigma\in\Sigma$ and all maximal strings $s,s'\in A$,
$s_\sfin=s'_\sfin$, where a string $s$ is maximal in $A$ if it is not a
generalized prefix of any other string in $A$. A set $A\subseteq
\Sigma^*$ is \emph{sparsely functional} if it is functional, and
moreover no string in $A$ is a generalized prefix of any other string
in $A$. Let $\cF$ be the set of sparsely functional subsets of
$\Sigma^*$, along with the empty set. Given $A$ a functional set, let
$\ulcorner A\urcorner$ be the largest set $A'\subseteq A$ that is
sparsely functional. The sparsely functional subsets of $\Sigma^*$
form a partially additive Kleene algebra $(\cF,+,\cdot,*,0,1)$ where
we take $A\summ B$ if $A\cup B$ is functional, in which case
$A+B=\ulcorner A\cup B\urcorner$, $A\cdot B=\ulcorner\{s\otimes s'\mid
s\in A, s'\in B\}\urcorner$, and $A^*=\ulcorner\Sigma\cup\{s\mid
\exists n(s\in A^n,s_\sfin\in\Omega)\}\urcorner$, where $A^0=\Omega$,
and $A^{n+1}=A\cdot A^n$. Thus, if we interpret $A$ as a partial
function from the initial states of the strings in $A$ to the final
states of these strings, then $A+B$ represents the union of partial
functions, $A\cdot B$ represents the composition of partial functions,
and $A^*$ represents the iterated application of $A$ until an element
of $\Omega$ is reached. The constant $0$ is taken to be the empty set,
and the constant $1$ is taken to be $\Sigma$.
\end{example}

A homomorphism of partially additive Kleene algebras is a function
$f:\cK\rightarrow \cK'$ such that if $a\summ b$ in $\cK$, then
$f(a)\summ f(b)$ in $\cK'$ and $f(a+b)=f(a)+f(b)$, as well as
$f(a\cdot b)=f(a)\cdot f(b)$, $f(a^*)=f(a)^*$, $f(0)=0$, and $f(1)=1$.
Partially additive Kleene algebras with homomorphisms form a category
$\cat{PKA}$. It follows from Example~\ref{x:total} that the category
$\cat{KA}$ of Kleene algebra forms a full subcategory of $\cat{PKA}$,
via the inclusion functor $\fP:\cat{KA}\rightarrow\cat{PKA}$.

Note that every partially additive Kleene algebra embeds in a total
Kleene algebra in a straightforward way, by adding a single element
$\star$. More precisely, let $\cK$ be a partially additive Kleene
algebra, and consider the Kleene algebra $\cK\cup\{\star\}$ with the
same operations, extended so that if $a$ and $b$ are not summable in
$\cK$, then $a+b=\star$ in $\cK\cup\{\star\}$, with $a+\star=\star$,
$0\cdot\star=\star\cdot 0=0$, $a\cdot \star=\star\cdot a=\star$, and
$\star^*=\star$. It is easy to check that $\cK\cup\{\star\}$ is a
Kleene algebra with a total operation $+$. Unfortunately, this
embedding is not universal. Intuitively, this construction does not
embed a partially additive Kleene algebra in the ``closest'' Kleene
algebra that contains it. We will see a universal embedding in the
next section.

\section{Star-Continuity}\label{s:pka-star}

A particularly interesting class of partially additive Kleene algebras
is one where the Kleene star operator is related to suprema with 
respect to $\le$. 
A partially additive Kleene algebra is
\emph{star-continuous} if 
$ab^{*}c=\sum_{n<\omega}ab^{n}c$, where $\sum$ is the supremum with
respect to the order $\le$.  This equation is equivalent to the
infinite implication
\begin{gather}
\text{if $ab^nc\summ w$ for all $n$ and
$\bigwedge_{n<\omega}ab^{n}c\le w$ then $ab^{*}c\le w$.} \label{e:star2}
\end{gather}
Equation \eqref{e:star2} does not always hold in partially additive
Kleene algebras, simply because not all Kleene algebras are
star-continuous \cite{r:kozen90}.  However, most natural examples of
partially additive Kleene algebra are in fact star-continuous. In
particular, Examples~\ref{x:monoid} and \ref{x:pfn} of the last
section are star-continuous.

\ifConf
\else

\subsection{Construction}

Star-continuous partially additive Kleene algebras form a full
subcategory $\cat{PKA}^*$ of $\cat{PKA}$. We now show that the
inclusion functor $\fK:\cat{PKA}^*\rightarrow\cat{PKA}$ witnessing the
fact that $\cat{PKA}^*$ is a subcategory has a left adjoint $S$ that
universally maps a partially additive Kleene algebra to the
``closest'' star-continuous partially additive Kleene algebra. The
idea is to take a partially additive Kleene algebra and impose
equation \eqref{e:star2} upon it. (The idea is similar to the
Abelianization of groups, which universally maps a group to an Abelian
group.) Note that there is nothing specific to partially additive
Kleene algebra here; this construction equally well maps a Kleene
algebra to the closest star-continuous Kleene algebra, and forms a
left adjoint to the appropriate inclusion functor.

\newcommand{\eqcl}[1]{[#1]_{\mathord{\equiv}}}

Constructing the left adjoint to $\fK$ can be done as follows. Let
$\cK$ be a partially additive Kleene algebra. Define $\equiv$ to be
the least congruence relation on $\cK$ such that for all $a,b,c,w$,
\begin{equation}
\label{e:cong}
\text{if $ab^{n}c+w\equiv w$ (for all $n\ge 0$) then
$ab^{*}c+w\equiv w$.}
\end{equation}
(In the presence of partiality, congruence for $+$ says that if
$x\summ z$, $y\summ z$, and $x\equiv y$, then $x+z\equiv y+z$.)  Let
$\eqcl{x}=\{y\mid x\equiv y\}$ be the equivalence class of $x$ under
congruence $\equiv$.  Let $\cK/\mathord{\equiv}$ be the set of
equivalence classes of $\equiv$. If we extend the operations of $\cK$
to equivalence classes in the standard way, it is easily seen that
$\cK/\mathord{\equiv}$ is a partially additive Kleene algebra that is
moreover star-continuous. (Say $\eqcl{x}\summ\eqcl{y}$ if there exists
$x'\equiv x$ and $y'\equiv y$ such that $x'\summ y'$, and set
$\eqcl{x}+\eqcl{y}=\eqcl{x'+y'}$.) If $f:\cK\rightarrow \cK'$ is a
partially additive Kleene algebra homomorphism, we can lift it to a
homomorphism $\hat{f}:\cK/\mathord{\equiv}\rightarrow
\cK'/\mathord{\equiv}$ by taking $\hat{f}(\eqcl{x})=\eqcl{f(x)}$.
\begin{lemma}\label{l:hatf}
If $f$ is a partially additive Kleene algebra homomorphism, then
$\hat{f}$ is a partially additive Kleene algebra homomorphism.
\end{lemma}
\begin{proof}
The main thing we need to check is that $\hat{f}$ is well-defined,
that is, if $x\equiv y$, then $\hat{f}(\eqcl{x})=\hat{f}(\eqcl{y})$,
that is, $f(x)\equiv f(y)$. Let $h:\cK' \rightarrow
\cK'/\mathord{\equiv}$ be the canonical surjective homomorphism
mapping an element $x$ of $\cK'$ to the equivalence class $\eqcl{x}$
of $\cK'/\mathord{\equiv}$. Recall that the kernel $\ker(g)$ of a
homomorphism $g:\cK\rightarrow\cK'$ is the set of all pairs of
elements $(x,y)$ such that $g(x)=g(y)$. We show that $\equiv\subseteq
\ker(h\circ f)$, so that $x\equiv y$ implies that $h(f(x))=h(f(y))$,
that is, $\eqcl{f(x)}=\eqcl{f(y)}$, as required. Recall that the
kernel of a homomorphism is a congruence; this holds even in the
presence of partiality for $+$: if $x\summ z$, $y\summ z$, and
$(x,y)\in\ker{f}$, then $f(x+z)=f(x)+f(z)=f(y)+f(z)=f(y+z)$, so that
$(x+z,y+z)\in\ker{f}$. Therefore, $\ker(h\circ f)$ is a congruence. It
therefore suffices to show that $\ker(h\circ f)$ is closed under
\eqref{e:star2} to get $\equiv\subseteq\ker{h\circ f}$. Assume that
for all $n$, $ab^n+c\summ w$  and 
$(ab^nc+w,w)\in\ker(h\circ f)$. In other words, we have
$h(f(ab^nc+w))=h(f(w))$ for all $n$. Because $h\circ f$ is a
homomorphism, we get $h(f(a))h(f(b))^nh(f(c))+h(f(w))=h(f(w))$ for all 
$n$. Since $\cK'/\mathord{\equiv}$ is star-continuous,
$h(f(a))h(f(b))^*h(f(c))+h(f(w))=h(f(w))$ holds, and therefore,
$h(f(ab^*c+w))=h(f(w))$; that is, $(ab^*c+w,w)\in\ker(h\circ f)$, as
required. 

Verifying that $\hat{f}$ is a partially additive Kleene algebra
homomorphism is straightforward. For instance, if
$\eqcl{x}\summ\eqcl{y}$, then
$\hat{f}(\eqcl{x}+\eqcl{y})=\hat{f}(\eqcl{x+y})=\eqcl{f(x+y)}=\eqcl{f(x)+f(y)}=\eqcl{f(x)}+\eqcl{f(y)}=\hat{f}(\eqcl{x})+\hat{f}(\eqcl{y})$.
\end{proof}

Let $S:\cat{PKA}\rightarrow\cat{PKA}^*$ map $\cK$ to $\cK/\mathord{\equiv}$, and 
$f$ to $\hat{f}$. It is easy to check that $S$ is a
functor. 
\begin{theorem}\label{t:Kadjoint}
The functor $\fS$ is a left adjoint of the inclusion functor
$\fK:\cat{PKA}^*\rightarrow\cat{PKA}$, via the adjunction map
$\phi:\Hom(S\cK,\cK')\rightarrow\Hom(\cK,K\cK')$ given by $(\phi
f)(x)=f(\eqcl{x})$.
\end{theorem}
\begin{proof}
To show that $\fS$ is a left adjoint to $\fK$, we need to show that
the adjunction map $\phi$ is a natural isomorphism between
$\Hom(\fS\cK,\cK')$ and $\Hom(\cK,\fK\cK')$. 

We first show that $\phi$ is an isomorphism. The inverse of $\phi$ is
given by the map $\psi$ defined by $(\psi g)(\eqcl{x})=g(x)$. We
check that this function is well defined, that is, if $x\equiv y$,
then $g(x)=g(y)$. The argument is similar
to that in the proof of Lemma~\ref{l:hatf}. Clearly, it suffices to
show that $\equiv\subseteq\ker(g)$. Since the kernel of a homomorphism
is a congruence, it suffices to show that $\ker(g)$ is closed under
\eqref{e:star2}. Let $g:\cK\rightarrow\fK\cK'$, where $\cK'$ is a
star-continuous partially additive Kleene algebra. Assume that for all
$n\ge 0$, $ab^nc\summ w$ and $(ab^nc+w,w)\in\ker(g)$, that is,
$g(ab^nc+w)=g(w)$ for all $n\ge 0$. Since $g$ is a homomorphism, we
get $g(a)g(b)^ng(c)+g(w)=g(w)$ for all $n\ge 0$. Since $\cK'$ is
star-continuous (and $\fK$ is the forgetful functor), we have
$g(a)g(b)^*g(c)+g(w)=g(w)$, that is, $g(ab^*c+w)=g(w)$, and
$(ab^*c+w,w)\in\ker(g)$, establishing the result. It is easy to see
that $(\psi g)$ is a partially additive Kleene algebra
homomorphism. To see that $\phi$ and $\psi$ are inverses, first let
$g:\cK\rightarrow\fK\cK'$, and note that
\begin{align*}
(\phi \circ \psi)(g)(x) 
 & = (\phi(\psi(g)))(x)\\
 & = (\psi(g))(\eqcl{x})\\
 & = g(x).
\end{align*}
Second, let $f:\fS\cK\rightarrow\cK'$, and note that
\begin{align*}
(\psi\circ\phi)(f)(\eqcl{x})
 & = (\psi(\phi(f)))(\eqcl{x})\\
 & = (\phi f)(x)\\
 & = f(\eqcl{x}).
\end{align*}

To establish the naturality of the adjunction map $\phi$, we need to
show the commutativity of two diagrams. First, we need to show that
for all morphisms $h:\cK'\rightarrow\cK''$ in $\cat{PKA}^*$, the
following diagram commutes:
\begin{diagram}[midshaft]
\Hom(\fS\cK,\cK') & \rTo^{\phi} & \Hom(\cK,\fK\cK')\\
 \dTo^{\Hom(\fS\cK,h)} & & \dTo_{\Hom(\cK,\fK h)}\\
\Hom(\fS\cK,\cK'') & \rTo^{\phi} & \Hom(\cK,\fK\cK'')
\end{diagram}
where $\Hom(\cK,-)$ is a functor that maps an object $\cK'$ to the set
$\Hom(\cK,\cK')$, and a morphism $h:\cK'\rightarrow\cK''$ to a morphism
$\Hom(\cK,h):\Hom(\cK,\cK')\rightarrow\Hom(\cK,\cK'')$, given by
$\Hom(\cK,h)(f)=h\circ f$. To verify the commutativity of the diagram,
let $h:\cK'\rightarrow\cK''$, let $f:\fS\cK\rightarrow\cK'$, and let
$x\in\cK$:
\begin{align*}
(\Hom(\cK,\fK h)\circ \phi)(f)(x) 
 & = (\Hom(\cK,\fK h)(\phi(f)))(x)\\
 & = ((\fK h)\circ (\phi f))(x)\\
 & = ((\fK h)((\phi f)(x)))\\
 & = h(f(\eqcl{x}))
\end{align*}
(since $\fK h = h$) and
\begin{align*}
(\phi\circ \Hom(\fS\cK,h))(f)(x)
 & = (\phi(\Hom(\fS\cK,h)(f)))(x)\\
 & = (\phi(h\circ f))(x)\\
 & = (h\circ f)(\eqcl{x})\\
 & = h(f(\eqcl{x})).
\end{align*}

Second, we need to show that for all morphisms
$h:\cK''\rightarrow\cK'$ in $\cat{PKA}$, the following diagram
commutes:
\begin{diagram}[midshaft]
\Hom(\fS\cK',\cK) & \rTo^{\phi} & \Hom(\cK',\fK\cK)\\
 \dTo^{\Hom(\fS h,\cK)} & & \dTo_{\Hom(h,\fK\cK)}\\
\Hom(\fS\cK'',\cK) & \rTo^{\phi} & \Hom(\cK'',\fK\cK)
\end{diagram}
where $\Hom(-,\cK)$ is a functor that maps an object $\cK'$ to the set
$\Hom(\cK',\cK)$, and a morphism $h:\cK''\rightarrow\cK'$ to a morphism
$\Hom(h,\cK):\Hom(\cK',\cK)\rightarrow\Hom(\cK'',\cK)$, given by
$\Hom(h,\cK)(f)=f\circ h$. To verify the commutativity of the diagram,
let $h:\cK''\rightarrow\cK'$, let $f:\fS\cK'\rightarrow\cK$, and let
$x\in\cK''$:
\begin{align*}
(\Hom(h,\fK\cK)\circ\phi)(f)(x) 
 & = (\Hom(h,\fK\cK)(\phi f))(x)\\
 & = ((\phi f)\circ h)(x)\\
 & = (\phi f)(h(x))\\
 & = f(\eqcl{h(x)})
\end{align*}
and
\begin{align*}
(\phi\circ\Hom(\fS h,\cK))(f)(x)
 & = (\phi(\Hom(\fS h,\cK)(f)))(x)\\
 & = (\phi(f\circ (\fS h)))(x)\\
 & = (\phi(f\circ \hat{h}))(x)\\
 & = (f\circ \hat{h})(\eqcl{x})\\
 & = f(\hat{h}(\eqcl{x}))\\
 & = f(\eqcl{h(x)}).
\end{align*}
\end{proof}

\subsection{Completion}

\fi

More interesting is the relationship between star-continuous Kleene
algebras and star-continuous partially additive Kleene algebras.  In
one direction, every star-continuous Kleene algebra is a
star-continuous partially additive Kleene algebra. Therefore, the
inclusion functor $P$, when restricted to the full subcategory
$\cat{KA}^*$ of star-continuous Kleene algebra, gives rise to an
inclusion functor (also named $P$) from $\cat{KA}^*$ to $\cat{PKA}^*$.
In the other direction, every star-continuous partially
additive Kleene algebra $\cK$ can be \emph{completed} to a
star-continuous Kleene algebra $\fT\cK$ in a universal way. Formally,
we show how to define a functor $\fT$ from the category $\cat{PKA}^*$
to the category $\cat{KA}^*$ that is the left adjoint to the inclusion
functor $\fP$.

The construction is a form of ideal completion, and is based on a
construction of Conway \citeyear{r:conway71}, later formalized by
Kozen \citeyear{r:kozen90}. We follow that treatment rather closely. A
\emph{star-ideal} $A$ in a star-continuous partially additive Kleene
algebra $\cK$ is a subset $A\subseteq \cK$ satisfying:
\begin{enumerate}
\item $A$ is nonempty;
\item $A$ is closed under $+$ (for summable pairs);
\item $A$ is closed downward under $\le$;
\item if $ab^nc$ is in $A$ for all $n$, then $ab^*c$ is in $A$. 
\end{enumerate}

The \emph{star-ideal generated by $A$}, denoted $\<A\>$, is the
smallest ideal containing $A$. It is equivalently defined as the
intersection of all the star-ideals containing $A$. If
$A=\{a_1,\dots,a_k\}$, we often write $\<a_1,\dots,a_k\>$ for
$\<A\>$. In particular, if $A=\{a\}$, then we write $\<a\>$ for
$\<A\>$. It is easy to check that $\<a\>=\{x\mid x\le a\}$. An ideal
$I$ is finitely generated if $I=\<A\>$ for some finite set $A$, and
countably generated if $I=\<A\>$ for some countable set $A$.
\ifConf\else
We can construct the star-ideal generated by $A$ directly, using a
transfinite argument. This is a useful technique for proving results,
as we shall see. 
\begin{lemma}\label{l:characterization}
Let $\tau$ be the map $\tau(A)=A\oplus A \cup A^{\le} \cup A^{\ostar}$,
where 
\begin{align*}
A \oplus B & = \{ a+b\mid a\in A, b\in B, \text{$a,b$ summable}\}\\
A^{\le} & = \{y\mid \exists x\in A. y\le x\}\\
A^{\ostar} & = \{ ab^*c \mid ab^{n}c\in A, \text{for all $n\ge 0$}\},
\end{align*}
and define the transfinite sequence
\begin{align*}
\tau^0(A) & = A\\
\tau^{\alpha+1}(A) & = \tau(\tau^\alpha(A))\\
\tau^{\lambda}(A) & = \bigcup_{\alpha<\lambda}\tau^\alpha(A)\qquad\text{$\lambda$ a limit ordinal}.
\end{align*}
For all $A\subseteq \cK$, there exists an ordinal $\kappa$ such
that $\tau^{\kappa}(A)=\tau^{\kappa+1}(A)$, and
$\tau^{\kappa}=\<A\>$. 
\end{lemma}
\begin{proof}
It is easy to check that $A\subseteq\tau(A)$, and that $\tau$ is
monotone, that is, $\tau(A)\subseteq\tau(B)$ whenever $A\subseteq
B$. The result then follows immediately by the Knaster-Tarski theorem
\cite{r:tarski55}. 
\end{proof}
\fi

Take $\fT\cK$ to be the set of all finitely generated star-ideals
of $\cK$. We can impose a star-continuous Kleene algebra structure on
$\fT\cK$ as follows. 
First, define the following operation $\odot$ on subsets of $\cK$:
\[ A \odot B  = \{a\cdot b \mid a\in A, b\in B\}.\]
\ifConf\else
\begin{lemma}\label{l:dotcupprops}
The following properties hold: 
\begin{enumerate}
\item $\<A\odot B\>= \<\<A\>\odot B\> = \<A \odot\<B\>\>$;
\item $\<A\cup B\>=\<\<A\> \cup B\> = \<A \cup \<B\>$.
\end{enumerate}
\end{lemma}
\begin{proof}
We prove part (2) here; part (1) is essentially Lemma 3 in Kozen
\citeyear{r:kozen90}, and is proved similarly. Since $\cup $ is
commutative, it is sufficient to show that $\<A\cup B\>=\<\<A\>\cup
B\>$. Since $A\subseteq \<A\>\subseteq \<\<A\>\cup B\>$ and
$B\subseteq \<\<A\>\cup B\>$, we have $\<A \cup B\>\subseteq
\<\<A\>\cup B\>$. We show the reverse inclusion $\<\<A\>\cup
B\>\subseteq\<A\cup B\>$ by transfinite induction, via the
characterization of star-ideals in
Lemma~\ref{l:characterization}. More precisely, we show that for all
ordinals $\alpha$, $\tau^\alpha (\<A\>\cup B)\subseteq \<A\cup
B\>$. It is easy to see that since $A\subseteq\<A\cup B\>$,
$\<A\>\subseteq\<A\cup B\>$, and similarly, $B\subseteq \<A\cup B\>$,
so that $\tau^0(\<A\>\cup B)=\<A\>\cup B\subseteq\<A\cup B\>$. For the
inductive case, 
\begin{align*}
\tau^{\alpha+1}(\<A\>\cup B) 
 & = \tau(\tau^{\alpha}(\<A\>\cup B))\\
 & \subseteq \tau(\<A\cup B\>)\qquad\text{by monotonicity of $\tau$}\\
 & = \<A\cup B\> \qquad\text{by property of star-ideals}.
\end{align*}
For limit ordinals $\lambda$,
\begin{align*}
\tau^{\lambda}(\<A\>\cup B) 
 & = \bigcup_{\alpha<\lambda}\tau^\alpha(\<A\>\cup B)\\
 & \subseteq \bigcup_{\alpha<\lambda}\<A\cup B\>\\
 & = \<A\cup B\>.
\end{align*}
Thus, $\<\<A\>\cup B\>=\tau^*(\<A\>\cup B)\subseteq \<A\cup B\>$, as required.
\end{proof}
Clearly, Lemma~\ref{l:dotcupprops} extends to arbitrary finite products and unions. 
\fi
If $I$ and $J$ are finitely generated star-ideals, generated
respectively by $A=\{a_1,\dots,a_k\}$ and $B$, define
\begin{align*}
I+J & = \<A\cup B\> \\
I\cdot J & = \<A\odot B\> \\
I^* & = \<(a_1^*\dots a_k^*)^*\> \\
1 & = \<1\>\\
0 & = \<0\>.
\end{align*}
Note that the operation $+$ on star-ideals is total.  We need to check
that these operations are well-defined; there can be many different
generators for a star-ideal, and the above operations need to give the
same result no matter what generators we take.
\ifConf
\else
\begin{lemma}\label{l:Twelldef}
If $I=\<A_1\>=\<A_2\>$ and $J=\<B_1\>=\<B_2\>$, then:
\begin{enumerate}
\item $\<A_1\cup B_1\>=\<A_2\cup B_2\>$;
\item $\<A_1\odot B_1\>=\<A_2\odot B_2\>$;
\item If $A_1=\<a_1,\dots,a_k\>$ and $B_1=\<b_1,\dots,b_l\>$, then
$\<(a_1^*\dots a_k^*)^*\>=\<(b_1^*\dots b_l^*)^*\>$.
\end{enumerate}
\end{lemma}
\begin{proof}
For part (1), by Lemma~\ref{l:dotcupprops}(1), we have $\<A_1\cup B_1\>=\< \<A_1\>\cup\<B_1\>\>=\<\<A_2\>\cup\<B_2\>\>=\<A_2\cup B_2\>$. 

For part (2), by Lemma~\ref{l:dotcupprops}(2), we have $\<A_1\odot B_1\>=\< \<A_1\>\odot\<B_1\>\>=\<\<A_2\>\odot\<B_2\>\>=\<A_2\odot B_2\>$. 

For part (3), we first establish that for all $i$, $a_i\le (b_1^*\dots
b_l^*)^*$. This follows easily from the fact that $a_i\in\<(a_1^*\dots
a_k^*)^*\>$, and thus that $a_i\in\<(b_1^*\dots
b_l^*)^*\>$, and by principality, $a_i\le(b_1^*\dots b_l^*)^*$.  The
result now follows almost immediately. For each $i$, by monotonicity
of $*$, $a_i^*\le ((b_1^*\dots b_l^*)^*)^*= (b_1^*\dots
b_l^*)^*$. Thus, $a_1^*\dots a_k^*\le (b_1^*\dots b_l^*)^*\dots
(b_1^*\dots b_l^*)^*\le (b_1^*\dots b_l^*)^*$. 
By monotonicity, $(a_1^*\dots a_k^*)^* \le ((b_1^*\dots b_l^*)^*)^* =
(b_1^*\dots b_l^*)^*$. Therefore, $\<(a_1^*\dots a_k^*)^*\>\subseteq
\<(b_1^*\dots b_l^*)^*\>$. A symmetric argument gives us the reverse
inclusion, establishing equality.
\end{proof}

We can check that $\le$ for star-ideals is just subset inclusion. 

\begin{lemma}\label{l:total-KA}
If $\cK$ is a star-continuous partially additive Kleene algebra, then
$\fT\cK$ is a star-continuous Kleene algebra.
\end{lemma}
\begin{proof}
Most of axioms of Kleene algebra (given by Kozen \citeyear{r:kozen94})
are straightforward to verify. We give a proof for the nontrivial
ones: $1+I\cdot I^*\le I^*$, $J+I\cdot K\le K$ implies $I^*\cdot J\le
K$, and the star-continuity condition. 

To show that $1+I\cdot I^*\le I^*$, it suffices to show
that $1\le I^*$ and $I\cdot I^*\le I^*$. Assume $I=\<A\>$, where
$A=\{a_1,\dots,a_k\}$.  First, since $\cK$ is a partially additive
Kleene algebra, $1\le (a_1^*\dots a_k^*)^*$, and so $1\in
\<(a_1^*\dots a_k^*)^*\>$, and $\<1\>\subseteq\<(a_1^*\dots
a_k^*)^*\>$, as required. Second, pick an arbitrary $a\in A$. Clearly,
$a\le a_1^*\dots a_k^*$, so $a(a_1^*\dots a_k^*)^* \le a_1^*\dots
a_k^* (a_1^*\dots a_k^*)^*
\le (a_1^*\dots a_k^*)^*$. Therefore, by closure of $I^*$,
$a(a_1^*\dots a_k^*)^*\in I^*$, for any $a\in A$. Thus, $A\odot
\{(a_1^*\dots a_k^*)^*\}\subseteq I^*$, and $\<A\odot\{(a_1^*\dots
a_k^*)^*\}\>\subseteq I^*$. But this is just $I\cdot I^*\le I^*$, as
required.

Next, we want to show that if $J+I\cdot K\le K$, then $I^*\cdot J\leq
K$. Assume $I=\<A\>$ (with $A=\{a_1,\dots,a_k\}$), $J=\<B\>$, and
$K=\<C\>$. Furthermore, assume $J+I\cdot K\le K$, that is, $\<B
\cup (A\odot C)\>\subseteq \<C\>$. By Lemma~\ref{l:dotcupprops}, this
is equivalent to 
\begin{equation}\label{e:sub}
\<\<B\>\cup (\<A\>\odot \<C\>)\>\subseteq \<C\>.
\end{equation}
We want to show that $\<\{(a_1^*\cdots a_k^*)^*\}\odot B\>\subseteq
\<C\>$. It is sufficient to show that for all $b\in B$, $(a_1^*\cdots
a_k^*)^*b\in\<C\>$. We know $b\in \<B\>$, thus by \eqref{e:sub},
$b\in\<C\>$. since $a_k\in\<A\>$, $a_k b\in\<A\>\odot\<C\>$, and by
\eqref{e:sub}, $a_k b\in\<C\>$. A straightforward induction shows that 
for all $n$, $a_k^n b\in\<C\>$. Therefore, by the closure properties
of $\<C\>$, $a_k^* b\in\<C\>$. A similarly argument with $a_k^* b$
instead of $b$ and $a_{k-1}$ instead of $a_k$ shows that $a_{k-1}^*
a_k^* b\in\<C\>$. A straightforward induction shows that $a_1^*\cdots
a_k^* b\in\<C\>$. Repeating the whole argument with $a_1^*\cdots a_k^* 
b$ instead of $b$ yields $(a_1^*\cdots a_k^*)^2 b\in\<C\>$. A
straightforward induction shows that for all $n$, $(a_1^*\cdots
a_k^*)^n b\in\<C\>$. By the closure properties of $\<C\>$,
$(a_1^*\cdots a_k^*)^* b\in\<C\>$, as required.

For the star-continuity condition, we want to show that if $I\cdot
J^n\cdot K\le L$ for all $n$, then $I\cdot J^*\cdot K\le L$. Assume
$I=\<A\>$, $J=\<B\>$ (where $B=\{b_1,\dots,b_k\}$), and
$K=\<C\>$. Furthermore, assume $I\cdot J^n\cdot K\le L$ for all $n$,
that is, $\<A\odot B^n \odot C\>\subseteq L$, where
$B^n=B\odot\dots\odot B$ ($n$ times). By Lemma~\ref{l:dotcupprops},
this is equivalent to
\begin{equation}\label{e:sub2}
\<\<A\>\odot \<B\>^n\odot \<C\>\>\subseteq L.
\end{equation}
It is sufficient to show that $A\odot\{(b_1^*\cdots b_k^*)^*\}\odot
C\subseteq L$. In other words, for all $a\in A$ and $c\in C$,
$a(b_1^*\cdots b_k^*)^*c\in L$. By the closure properties of $L$, it
is sufficient to show that for all $a\in A$, all $c\in C$, and all
$m\ge 0$, $a(b_1^*\dots b_k^*)^m c\in L$. For $m=0$, the result
follows trivially from \eqref{e:sub2} (taking $n=0$). For $m>0$, by
\eqref{e:sub2}, for all $n_1^1,\dots,n_k^1,\dots,n_1^m,\dots,n_k^m$,
we have $a (b_1^{n_1^1}\cdots b_k^{n_k^1})\cdots(b_1^{n_1^m}\cdots
b_k^{n_k^m})c\in L$. By the closure properties of $L$, we have \[a
(b_1^* b_2^{n_2^1}\cdots b_k^{n_k^1})\cdots(b_1^{n_1^m}\cdots
b_k^{n_k^m})c\in L.\] An easy induction shows that $a 
(b_1^* \cdots b_k^*)\cdots (b_1^*\cdots b_k^*)c\in L$, that is,
$a(b_1^*\cdots b_k^*)^m c\in L$, as required.
\end{proof}
\fi

We can extend $\fT$ to a functor by specifying its action on
$\cat{PKA}^*$ morphisms. If $f:\cK\rightarrow\cK'$ is a homomorphism
of partially additive Kleene algebras, define $\fT
f:\fT\cK\rightarrow\fT\cK'$ as $(\fT f)(I)=\<f[I]\>$, where
$f[A]=\{f(a)\mid a\in A\}$. 
\ifConf
\else
\begin{lemma}\label{l:fstarideal}
$\<f[\<A\>]\>=\<f[A]\>$.
\end{lemma}
\begin{proof}
This is a straightforward adaption of the proof of Lemma 4 in Kozen
\citeyear{r:kozen90}. 
\end{proof}

\begin{lemma}\label{l:fthom}
$\fT f$ is a Kleene algebra homomorphism.
\end{lemma}
\begin{proof}
\begin{align*}
(\fT f)(\<0\>) 
 & = \< f[\<0\>]\>\\
 & = \< f(0) \> \qquad\text{by Lemma~\ref{l:fstarideal}}\\
 & = \< 0\>
\end{align*}

\begin{align*}
(\fT f)(\<1\>) 
 & = \< f[\<1\>]\>\\
 & = \< f[1] \> \qquad\text{by Lemma~\ref{l:fstarideal}}\\
 & = \< 1 \>.
\end{align*}

Assume $I=\<A\>$ and $J=\<B\>$,
\begin{align*}
(\fT f)(I+J)
 & = \< f[I+J] \>\\
 & = \< f[\<A\cup B\>] \>\\
 & = \< f[A\cup B]\> \qquad\text{by Lemma~\ref{l:fstarideal}}\\
 & = \< f[A]\cup f[B]\> \\
 & = \<f[A]\>+\<f[B]\>\\
 & = \<f[\<A\>]\> + \<f[\<B\>]\>\qquad\text{by Lemma~\ref{l:fstarideal}}\\
 & = \<f[I]\> + \<f[J]\>\\
 & = (\fT f)(I) + (\fT f)(J)
\end{align*}
\begin{align*}
(\fT f)(I\cdot J) 
 & = \<f[I\cdot J]\>\\
 & = \<f[\<A \odot B\>]\>\\
 & = \<f[A \odot B]\>  \qquad\text{by Lemma~\ref{l:fstarideal}}\\
 & = \< \{ f(a\cdot b)\mid a\in A,b\in B\} \>\\
 & = \< \{ f(a)\cdot f(b)\mid a\in A,b\in B\}\>\\
 & = \< \{ x\cdot y\mid x\in f[A],y\in f[B]\}\>\\
 & = \< f[A]\odot f[B]\>\\
 & = \< f[A]\> \cdot \<f[B]\>\\
 & = \< f[\<A\>]\>\cdot\<f[\<B\>]\>\qquad\text{by Lemma~\ref{l:fstarideal}}\\
 & = \< f[I]\>\cdot \<f[J]\>  = (\fT f)(I)\cdot (\fT f)(J).
\end{align*}

Assume $I=\<\{a_1,\dots,a_k\}\>$,
\begin{align*}
(\fT f)(I^*) 
 & = \<f [I^*]\>\\
 & = \<f [\<\{(a_1^*\cdots a_k^*)^*\}\>]\>\qquad\text{by Lemma~\ref{l:fstarideal}}\\
 & = \<f [\{ (a_1^*\cdots a_k^*)^*\}]\>\\
 & = \< \{ f ((a_1^*\cdots a_k^*)^*)\}\>\\
 & = \< \{ (f(a_1)^*\cdots f(a_k)^*)^*\} \>\\
 & = (\<\{f(a_1),\dots,f(a_k)\}\>)^*\\
 & = (\<f[\{a_1,\dots,a_k\}]\>)^*\\
 & = (\<f[\<\{a_1,\dots,a_k\}\>]\>)^*\qquad\text{by Lemma~\ref{l:fstarideal}}\\
 & = (\<f[I]\>)^*  = ((\fT f)(I))^*.
\end{align*}
\end{proof}
It is straightforward to check that $T$ is a functor.
\fi

\begin{theorem}\label{t:Padjoint}
The functor $\fT$ is a left adjoint to the inclusion functor
$\fP:\cat{KA}^*\rightarrow\cat{PKA}^*$, via the adjunction map
$\phi:\Hom(\fT\cK,\cK')\rightarrow\Hom(\cK,\fP\cK')$ given by
$(\phi f)(x)=f(\<x\>)$.
\end{theorem}
\ifConf\else
\begin{proof}
To show that $\fT$ is a left adjoint to $\fP$, we need to show that
the adjunction map $\phi$ is a natural isomorphism between
$\Hom(\fT\cK,\cK')$ and $\Hom(\cK,\fP\cK')$. 

We first show that $(\phi f)$ is a partially additive Kleene algebra
homomorphism. The verification is mostly straightforward. 
\begin{align*}
(\phi f)(0) 
 & = f(\<0\>)\\
 & = \<0\>,
\end{align*}
and similarly for $1$.
\begin{align*}
(\phi f)(x+y) 
 & = f(\<x+y\>)\\
 & = f(\<\{x,y\}\>)\\
 & = f(\<x\>+\<y\>)\\
 & = f(\<x\>)+f(\<y\>)\\
 & = (\phi f)(x) + (\phi f)(y)
\end{align*}
\begin{align*}
(\phi f)(x\cdot y)
 & = f(\<x\cdot y\>)\\
 & = f(\<\{x\}\odot\{y\}\>)\\
 & = f(\<x\>\cdot\<y\>)\\
 & = f(\<x\>)\cdot f(\<y\>)\\
 & = (\phi f)(x) \cdot (\phi f)(y)
\end{align*}
\begin{align*}
(\phi f)(x^*)
 & = f(\<x^*\>)\\
 & = f(\<(x^*)^*\>)\\
 & = f(\<x\>^*)\\
 & = (f(\<x\>))^*\\
 & = ((\phi f)(x))^*.
\end{align*}

We next show that $\phi$ is an isomorphism. The inverse of $\phi$ is
given by the map $\psi$ defined by $(\psi g)(I)=g(a_1)+\dots+g(a_k)$
if $I=\<a_1,\dots,a_k\>$.  We check that this function is well
defined, that is, if $I=\<a_1,\dots,a_k\>$ and $I=\<b_1,\dots,b_l\>$,
then $g(a_1)+\dots+g(a_k)=g(b_1)+\dots+g(b_l)$. First, since
$g:\cK\rightarrow\fP\cK'$ and since $\fP$ is the forgetful functor from
$\cat{KA}^*$, the operation $+$ is total on the range of $g$. Note
that from Lemma~\ref{l:fstarideal}, we have that
$\<g[I]\>=\<g(a_1),\dots,g(a_k)\>=\<g(b_1),\dots,g(b_l)\>\>\subseteq
\<g(b_1)+\dots+g(b_l)\> = (g(b_1)+\dots+g(b_l))^{\le}$. Therefore,
$g(a_i)\le g(b_1)+\dots+g(b_l)$ for all $i$, and therefore
$g(a_1)+\dots+g(a_k)\le g(b_1)+\dots+g(b_l)$. A symmetric argument
shows that $g(b_1)+\dots+g(b_l)\le g(a_1)+\dots+g(a_k)$, establishing
the result.  We check that $(\psi g)$ is a Kleene algebra
homomorphism.
\begin{align*}
(\psi g)(\<0\>) 
 & =   g(0)\\
 & = 0
\end{align*}
and similarly for $1$.
\begin{align*}
(\psi g)(\<A\>+\<B\>) 
 & = (\psi g)(\<A\cup B\>)\\
 & = \sum g[A\cup B]\\
 & = \sum g[A]\cup g[B]\\
 & = \sum g[A] + \sum g[B]\\
 & = (\psi g)(\<A\>) + (\psi g)(\<B\>)
\end{align*}
\begin{align*}
(\psi g)(\<A\>\cdot\<B\>) 
 & = (\psi g)(\<A\odot B\>)\\
 & = \sum g[A\odot B]\\
 & = \sum g[A]\odot g[B]\\
 & = \sum g[A] \cdot \sum g[B]\\
 & = (\psi g)(\<A\>)\cdot(\psi g)(\<B\>)
\end{align*}
\begin{align*}
(\psi g)(\<\{a_1,\dots,a_k\}\>^*) 
 & = (\psi g)(\<(a_1^*\cdots a_k^*)^*\>)\\
 & = g((a_1^*\cdots a_k^*)^*)\\
 & = (g(a_1)^*\cdots g(a_k)^*)^*\\
 & = (g(a_1)+\dots+g(a_k))^*\\
 & = (\sum g[\{a_1,\dots,a_k\}])^*\\
 & = ((\psi g)(\<a_1,\dots,a_k\>))^*.
\end{align*}
To see that $\phi$ and $\psi$ are inverses, first let
$g:\cK\rightarrow\fP\cK'$, and note that
\begin{align*}
(\phi\circ \psi)(g)(x)
 & = (\phi(\psi(g)))(x)\\
 & = (\psi g)(\<x\>)\\
 & = g(x).
\end{align*}
Similarly, let $f:\fT\cK\rightarrow\cK'$, and note that
\begin{align*}
(\psi\circ\phi)(f)(\<\{a_1,\dots,a_k\}\>)
 & = (\psi(\phi(f)))(\<\{a_1,\dots,a_k\}\>)\\
 & = (\phi f)(a_1)+\dots+(\phi f)(a_k)\\
 & = f(\<a_1\>) + \dots f(\<a_k\>)\\
 & = f(\<a_1\>+\dots+\<a_k\>)\\
 & = f(\<a_1,\dots,a_k\>).
\end{align*}

To establish the naturality of the adjunction map $\phi$, we need to
show the commutativity of two diagrams. (See the proof of
Theorem~\ref{t:Kadjoint}.) First, let $h:\cK'\rightarrow\cK''$, let
$f:\fT\cK\rightarrow\cK'$, and let 
$x\in\cK$:
\begin{align*}
(\Hom(\cK,\fP h)\circ \phi)(f)(x) 
 & = (\Hom(\cK,\fP h)(\phi(f)))(x)\\
 & = ((\fP h)\circ (\phi f))(x)\\
 & = ((\fP h)((\phi f)(x)))\\
 & = h(f(\<x\>))
\end{align*}
(since $\fP h = h$) and
\begin{align*}
(\phi\circ \Hom(\fT\cK,h))(f)(x)
 & = (\phi(\Hom(\fT\cK,h)(f)))(x)\\
 & = (\phi(h\circ f))(x)\\
 & = (h\circ f)(\<x\>)\\
 & = h(f(\<x\>)).
\end{align*}

Second, let $h:\cK''\rightarrow\cK'$, let $f:\fT\cK'\rightarrow\cK$,
and let $x\in\cK''$:
\begin{align*}
(\Hom(h,\fP\cK)\circ\phi)(f)(x) 
 & = (\Hom(h,\fP\cK)(\phi f))(x)\\
 & = ((\phi f)\circ h)(x)\\
 & = (\phi f)(h(x))\\
 & = f(\<h(x)\>)
\end{align*}
and
\begin{align*}
(\phi\circ\Hom(\fT h,\cK))(f)(x)
 & = (\phi(\Hom(\fT h,\cK)(f)))(x)\\
 & = (\phi(f\circ (\fT h)))(x)\\
 & = (f\circ (\fT h))(\<x\>)\\
 & = f((\fT h)(\<x\>))\\
 & = f(\<h(x)\>).
\end{align*}
\end{proof}
\fi

\section{Relationship with Partially Additive Idempotent Semirings}\label{s:ps} 
An algebraic structure that is quite similar to partially additive Kleene
algebras was studied by Arbib and Manes \citeyear{r:manes86}, who introduced
a form of partially additive semiring with an infinitary sum
operation.\footnote{Strictly speaking, Arbib and Manes define the
notion of a \emph{partially additive category}. A partially additive
semiring is a one-object partially additive category.}
At the basis of this structure is the
notion of a \emph{partially additive monoid}. Roughly speaking, a
partially additive monoid is like a monoid (that is, a set with a
single associate operation and an identity element), except that the
operation is only partially defined, and is infinitary. In this
section, we show the relationship between partially additive Kleene
algebras and partially additive idempotent semirings.

To make this precise, we need some terminology. 
Let $\cM$ be a fixed set. An \emph{$\cI$-indexed family} in $\cM$ is a
function $x:\cI\longrightarrow \cM$, written $(x_i\mid i\in \cI)$. We
usually write $x_i$ rather than $x(i)$. The (necessarily unique)
$\emptyset$-indexed family is called the \emph{empty family}.  A
family $(y_j \mid j\in \cJ)$ is said to be a
\emph{subfamily} of $(x_i\mid i\in \cI)$ if $\cJ\subseteq \cI$ and $y_j=x_j$
for all $j\in \cJ$. A family $(x_i\mid i\in \cI)$ is \emph{countable} if
$\cI$ is a countable set. Given an index set $\cI$, the family $(\cI_j\mid
j\in \cJ)$ is a \emph{partition} of $\cI$ if $\cI_j\cap \cI_k=\empty$ whenever
$j\ne k$, and $\cI=\cup_{j\in \cJ} \cI_j$. Note that we allow an element
$\cI_j$ of the partition to be empty.

A \emph{partially additive monoid} is a pair $(\cM,\sum)$ where $\cM$ is 
a nonempty set and $\sum$ is a partial function mapping countable
families in $\cM$ to elements of $\cM$ (the family $(x_i\mid i\in \cI)$ is
\emph{summable} if $\sum(x_i\mid i\in \cI)$ is defined) subject to
the following three conditions: 
\begin{enumerate}

\item \textit{Partition-associativity axiom:} If $(x_i\mid i\in \cI)$ is a
countable family and $(\cI_j\mid j\in \cJ)$ is a partition of $\cI$ with
$\cJ$ countable, then $(x_i\mid i\in \cI)$ is summable if and only if it 
the case both that for every $j\in \cJ$, $(x_i \mid i\in \cI_j)$ is
summable, and $(\sum(x_i\mid i\in \cI_j)\mid j\in \cJ)$ is
summable. In that case, $\sum(x_i\mid i\in
\cI)=\sum(\sum(x_i\mid i\in \cI_j)\mid j\in \cJ)$;

\item \textit{Unary sum axiom:} Any family $(x_i\mid i\in \cI)$ where $|\cI|=1$
is summable, and $\sum(x_i\mid i\in \cI)=x_j$ if $\cI=\{j\}$;

\item \textit{Limit axiom:} If $(x_i\mid i\in \cI)$ is a countable family,
and if the subfamily $(x_i\mid i\in F)$ is summable for every finite 
set $F\subseteq \cI$, then $(x_i\mid i\in \cI)$ is summable.
\end{enumerate}

One consequence of the partition-associativity axiom is that every
partially additive monoid is commutative: the order of the elements of
the family being summed does not affect the sum. Moreover, in every
partially additive monoid there exists a special element that acts as
the identity under sums. More precisely, for a partially additive
monoid $\cM$, let $!:\emptyset\longrightarrow \cM$ be the empty family in
$\cM$, which is summable by an application of the
partition-associativity axiom. Let $0$ be $\sum !$. That this $0$ acts
as the identity is captured by the following easily-proved fact: if
$(x_i\mid i\in \cI)$ is a summable family in $\cM$, $\cJ$ a countable set
disjoint from $\cI$, and $x_i=0$ for $i\in \cJ$, then $(x_i\mid i\in \cI\cup
\cJ)$ is summable, and $\sum(x_i\mid i\in \cI\cup \cJ)=\sum(x_i\mid i\in
\cI)$. 

A \emph{partially additive idempotent semiring} is a structure
$(\cS,\sum,\cdot,1)$
where 
\begin{enumerate}
\item $(\cS,\sum)$ is a partially additive monoid that is idempotent:
if $\cI$ is a countable set and we form the family $(x_i\mid i\in
\cI)$ with $x_i=x$ for all $i\in \cI$, then $(x_i\mid i\in \cI)$ is
summable, and $\sum(x_i\mid i\in \cI)=x$;
\item $(\cS,\cdot,1)$ is a monoid;
\item If $(x_i\mid i\in\cI)$ is summable, then $(y\cdot x_i\mid
i\in\cI)$ and $(x_i\cdot z\mid i\in\cI)$ are summable, and
$y(\sum(x_i\mid i\in\cI))=\sum(y\cdot x_i\mid i\in\cI)$ and
$(\sum(x_i\mid i\in\cI))z = \sum(x_i\cdot z\mid i\in\cI)$;
\item $a\cdot 0=0\cdot a = 0$.
\end{enumerate}
Partially additive idempotent semirings form a category $\cat{PS}$,
with morphisms being homomorphisms that preserve the sum of summable
countable families. (These homomorphisms are sometimes called
$\omega$-continuous.) 

In the examples treated in the literature, the infinitary sum
is used almost exclusively to define the $*$ operation, by taking  
$b^* = \sum(b^n\mid n\in\bbN)$,
where $b^0=1$ and $b^{n+1}=b b^n$. By countable
distributivity, we have
$ab^*c = \sum(ab^nc\mid n\in\bbN)$,
and therefore every partially additive idempotent semiring is a
star-continuous partially additive Kleene algebra. This gives us a
forgetful functor $U:\cat{PS}\rightarrow\cat{PKA}^*$.

There is a way to universal embed a star-continuous partially additive 
Kleene algebra into a partially additive idempotent semiring, by
essentially extending the $+$ operation to an infinitary
summation. This takes the form of a functor $\fC$ from $\cat{PKA}^*$
to $\cat{PS}$. It is again a form of ideal completion, of the kind
already seen in Section~\ref{s:pka-star}. We use the same definition
of star-ideals. This time, however, we take $\fC\cK$ to
be the set of \emph{countably generated} star-ideals of $\cK$. We
define the partially additive idempotent semiring operations as
follows. Most of the operations are defined as we did for the functor
$\fT$ in Section~\ref{s:pka-star}.
If $I$ and $J$ are countably generated star-ideals, generated
respectively by the countable sets $A$ and $B$, define
\begin{align*}
I\cdot J & = \<A\odot B\> \\
1 & = \<1\>\\
0 & = \<0\>.
\end{align*}
To define the infinitary summation $\sum$, let $(I_i\mid i\in \cI)$
be a countable family of countably generated star-ideals of $\cK$,
where $I_i$ is generated by the countable set $A_i$. Say that the
family $(I_i\mid i\in \cI)$ is summable if for all finite subsets
$\{a_1,\dots,a_k\}\subseteq \bigcup_{i\in\cI}A_i$, the finite sum $a_1+\dots+a_k$
is defined in $\cK$, for all possible ways of associating the binary
$+$ operations in $a_1+\dots+a_k$; if $(I_i\mid i\in \cI)$ is summable, define
$\sum(I_i\mid i\in \cI) = \<\bigcup_{i\in\cI} A_i\>$. Again, we can check that
these operations are well-defined, by showing that they give the same
star-ideals irrespectively of the choice of countable generating
sets. 
\ifConf\else(The details are as in Lemma~\ref{l:Twelldef}.)\fi

It is straightforward to check that $\fC\cK$ is a partially additive
idempotent semiring.  We can extend $\fC$ to a functor by specifying
its action on $\cat{PKA}^*$ morphisms. If $f:\cK\rightarrow\cK'$ is a
homomorphism of partially additive Kleene algebras, define $\fC
f:\fC\cK\rightarrow\fC\cK'$ as $(\fC f)(I)=\<f[I]\>$, where
$f[A]=\{f(a)\mid a\in A\}$.
\ifConf
\else
\begin{lemma}
$\fC f$ is a partially additive idempotent semiring homomorphism. 
\end{lemma}
\begin{proof}
This proof is similar to that of Lemma~\ref{l:fthom}. The only case
that is different (aside from the fact that there is no case for $*$,
since there is no such operation in partially additive idempotent
semirings) is the countable sum. 
Let $(I_i\mid i\in\cI)$ be a countable family of countably generated
star-ideals, where $I_i=\<A_i\>$. Assume $(I_i\mid i\in\cI)$ is
summable. 
\begin{align*}
(\fC f)(\sum(I_i\mid i\in\cI)) 
 & = \<f[\sum(I_i\mid i\in\cI)]\>\\
 & = \<f[\<\bigcup_{i\in\cI} A_i\>]\>\\
 & = \<f[\bigcup_{i\in\cI} A_i]\>\qquad\text{by Lemma~\ref{l:fstarideal}}\\
 & = \<\bigcup_{i\in\cI} f[A_i]\>\\
 & = \sum(\<f[A_i]\>\mid i\in\cI)\\
 & = \sum(\<f[\<A_i\>]\>\mid i\in\cI)\qquad\text{by Lemma~\ref{l:fstarideal}}\\
 & = \sum(\<f[I_i]\>\mid i\in\cI)\\
 & = \sum((\fC f)(I_i)\mid i\in\cI).
\end{align*}
\end{proof}
It is straightforward to check that $\fC$ is a functor.
\fi
\begin{theorem}\label{t:Uadjoint}
The functor $\fC$ is a left adjoint of the inclusion functor
$U:\cat{PS}\rightarrow\cat{PKA}^*$, via the adjunction map
$\phi:\Hom(\fC\cK,\cS)\rightarrow\Hom(\cK,\fU\cS)$ given by
$(\phi f)(x)=f(\<x\>)$.
\end{theorem}
\ifConf\else
\begin{proof}
To show that $\fC$ is a left adjoint to $\fU$, we need to show that
the adjunction map $\phi$ is a natural isomorphism between
$\Hom(\fC\cK,\cS)$ and $\Hom(\cK,\fU\cS)$. 

We first show that $(\phi f)$ is a partially additive Kleene algebra
homomorphism. The verification is mostly straightforward. 
\begin{align*}
(\phi f)(0) 
 & = f(\<0\>)\\
 & = \<0\>,
\end{align*}
and similarly for $1$.
\begin{align*}
(\phi f)(x+y) 
 & = f(\<x+y\>)\\
 & = f(\<\{x,y\}\>)\\
 & = f(\<x\>+\<y\>)\\
 & = f(\<x\>)+f(\<y\>)\\
 & = (\phi f)(x) + (\phi f)(y)
\end{align*}
\begin{align*}
(\phi f)(x\cdot y)
 & = f(\<x\cdot y\>)\\
 & = f(\<\{x\}\odot\{y\}\>)\\
 & = f(\<x\>\cdot\<y\>)\\
 & = f(\<x\>)\cdot f(\<y\>)\\
 & = (\phi f)(x) \cdot (\phi f)(y).
\end{align*}
For $x^*$, note that $\<x^*\>=\<\bigcup_{n\ge 0}\{x^n\}\>$: since
$x^n\le x^*$ for all $n\ge 0$, $\bigcup_{n\ge 0}\{x^n\}\subseteq
\<x^*\>$ by the closure properties of star-ideals, and thus
$\<\bigcup_{n\ge 0}\>\subseteq\<x^*\>$; conversely, we have
$x^n\in\<\bigcup_{n\ge 0}\{x^n\}$ for all $n\ge 0$, and again by the
closure properties of star-ideals, $x^*\in\<\bigcup_{n\ge
0}\{x^n\}\>$, and thus $\<x^*\>\subseteq\<\bigcup_{n\ge 0}\{x^n\}\>$. 
\begin{align*}
(\phi f)(x^*) 
 & = f(\<x^*\>)\\
 & = f(\<\bigcup_{n\ge 0}\{x^n\}\>)\\
 & = f(\sum(\<x^n\>\mid n\ge 0))\\
 & = f(\sum(\<x\>^n\mid n\ge 0))\\
 & = \sum(f(\<x\>)^n\mid n\ge 0))\\
 & = \sum((\phi f)(x)^n\mid n\ge 0)\\
 & = ((\phi f)(x))^* \qquad\text{by definition of $*$}. 
\end{align*}

We next show that $\phi$ is an isomorphism. The inverse of $\phi$ is
given by the map $\psi$ defined by $(\psi g)(I)=\sum(g(a)\mid a\in A)$
if $I=\<A\>$. We check that this function is well defined, that is, if
$I=\<A\>$ and $I=\<B\>$, then $\sum(g(a)\mid a\in A) = \sum(g(B)\mid
b\in B)$. The argument is essentially the same as that in the proof of
Theorem~\ref{t:Padjoint}. We repeat the argument here for
convenience. Note that from Lemma~\ref{l:fstarideal}, we have that
$\<g[I]\>=\<g[A]\>=\<g[B]\>\subseteq
\<\sum(g(b)\mid b\in B)\> = (\sum(g(b)\mid b\in B))^{\le}$. Therefore,
$g(a)\le \sum(g(b)\mid b\in B)$ for all $a\in A$, and therefore
$\sum(g(a)\mid a\in A)\le \sum(g(b)\mid b\in B)$. A symmetric argument
shows that $\sum(g(b)\mid b\in B)\le \sum(g(a)\mid a\in A)$,
establishing the result.  We check that $(\psi g)$ is a partially
additive idempotent semiring homomorphism.
\begin{align*}
(\psi g)(\<0\>) 
 & =   g(0)\\
 & = 0
\end{align*}
and similarly for $1$.
\begin{align*}
(\psi g)(\sum(\<A_i\>\mid i\in\cI))
 & = (\psi g)(\<\bigcup_{i\in\cI}A_i\>)\\
 & = \sum (g(a)\mid a\in\bigcup_{i\in\cI}A_i)\\
 & = \sum (\sum (g(a)\mid a\in A_i)\mid i\in\cI)\\
 & = \sum ((\psi g)(\<A_i\>)\mid i\in\cI)
\end{align*}
\begin{align*}
(\psi g)(\<A\>\cdot\<B\>) 
 & = (\psi g)(\<A\odot B\>)\\
 & = \sum g[A\odot B]\\
 & = \sum g[A]\odot g[B]\\
 & = \sum g[A] \cdot \sum g[B]\\
 & = (\psi g)(\<A\>)\cdot(\psi g)(\<B\>).
\end{align*}

To see that $\phi$ and $\psi$ are inverses, first let
$g:\cK\rightarrow\fU\cS$, and note that
\begin{align*}
(\phi\circ \psi)(g)(x)
 & = (\phi(\psi(g)))(x)\\
 & = (\psi g)(\<x\>)\\
 & = g(x).
\end{align*}
Similarly, let $f:\fC\cK\rightarrow\cS$, and note that
\begin{align*}
(\psi\circ\phi)(f)(\<A\>)
 & = (\psi(\phi(f)))(\<A\>)\\
 & = \sum ((\phi f)(a)\mid a\in A)\\
 & = \sum (f(\<a\>)\mid a\in A)\\
 & = f (\sum (\<a\>\mid a\in A))\\
 & = f (\<A\>).
\end{align*}

To establish the naturality of the adjunction map $\phi$, we need to
show the commutativity of two diagrams. (See the proof of
Theorem~\ref{t:Kadjoint}.) First, let $h:\cS\rightarrow\cS'$, let
$f:\fC\cK\rightarrow\cS$, and let 
$x\in\cK$:
\begin{align*}
(\Hom(\cK,\fU h)\circ \phi)(f)(x) 
 & = (\Hom(\cK,\fU h)(\phi(f)))(x)\\
 & = ((\fU h)\circ (\phi f))(x)\\
 & = ((\fU h)((\phi f)(x)))\\
 & = h(f(\<x\>))
\end{align*}
(since $\fU h = h$) and
\begin{align*}
(\phi\circ \Hom(\fC\cK,h))(f)(x)
 & = (\phi(\Hom(\fC\cK,h)(f)))(x)\\
 & = (\phi(h\circ f))(x)\\
 & = (h\circ f)(\<x\>)\\
 & = h(f(\<x\>)).
\end{align*}

Second, let $h:\cK'\rightarrow\cK$, let $f:\fC\cK\rightarrow\cS$,
and let $x\in\cK'$:
\begin{align*}
(\Hom(h,\fU\cS)\circ\phi)(f)(x) 
 & = (\Hom(h,\fU\cS)(\phi f))(x)\\
 & = ((\phi f)\circ h)(x)\\
 & = (\phi f)(h(x))\\
 & = f(\<h(x)\>)
\end{align*}
and
\begin{align*}
(\phi\circ\Hom(\fC h,\cS))(f)(x)
 & = (\phi(\Hom(\fC h,\cS)(f)))(x)\\
 & = (\phi(f\circ (\fC h)))(x)\\
 & = (f\circ (\fC h))(\<x\>)\\
 & = f((\fC h)(\<x\>))\\
 & = f(\<h(x)\>).
\end{align*}
\end{proof}
\fi

\section{Relationship with Closed Semirings}\label{s:cs}

\emph{Closed semirings} \cite{r:aho75,r:mehlhorn84} are often used as
alternatives to Kleene algebras to treat iteration by defining it via
an infinite sum operation. In fact, a closed semiring is simply a
partially additive idempotent semiring where the $\sum$ operator is
always defined, that is, where every countable family is summable.  In
this section, we show that there are two ways to universally map
star-continuous partially additive Kleene algebras to closed
semirings, using the constructions of the previous two sections; these
nevertheless yield isomorphic closed semirings.

In Section~\ref{s:pka-star}, we showed that star-continuous partially
additive Kleene algebras can be completed to star-continuous
Kleene algebras, via the functor $\fT$. Kozen
\citeyear{r:kozen90} showed that every star-continuous Kleene algebra can
be universally embedded into a closed semiring, via a functor
$\fCp:\cat{KA}^*\rightarrow\cat{CS}$ (where $\cat{CS}$ is the category 
of closed semirings) and that $\fCp$ is a left
adjoint to the inclusion functor $\fUp:\cat{CS}\rightarrow\cat{KA}^*$.
By composing the functors, we get the adjunctions
\begin{diagram}[midshaft]
\cat{PKA}^* &  \pile{\rTo^\fT\\ \lTo_\fP} & \cat{KA}^* &
\pile{\rTo^\fCp\\ \lTo_\fUp} & \cat{CS}.
\end{diagram}
Thus, the composite functor $\fCp\circ\fT$ is a left adjoint to the
forgetful functor $\cat{CS}\rightarrow\cat{PKA}^*$. 

In Section~\ref{s:ps} we showed that every star-continuous partially
additive Kleene algebra can be embedded into a partially additive
idempotent semirings, via the functor
$\fC:\cat{PKA}^*\rightarrow\cat{PS}$. 
\ifConf
\else
It is straightforward to show that partially
additive idempotent semirings can be completed to closed semirings in
a universal way; this construction is quite similar to the
completion of partially additive Kleene algebras. This construction is 
essentially the one described by Manes \citeyear{r:manes88}. If $\cS$ is a
partially additive idempotent semiring, define an \emph{ideal} of
$\cS$ to be a subset $A\subseteq\cS$ satisfying:
\begin{enumerate}
\item $A$ is nonempty;
\item $A$ is closed under $\sum$ (for summable families);
\item $A$ is closed downward under $\le$.
\end{enumerate}
We use the same notation as for star-ideals, and write $\<A\>$ for the
smallest ideal generated by the set $A$. The analogue of results for
star-ideals carry over (for instance, Lemmas~\ref{l:characterization}
and \ref{l:dotcupprops}). If $\cS$ is a partially additive idempotent
semiring, let $\fTp\cS$ be the set of ideals of $\cS$.  We can impose a
closed semiring structure on $\fTp\cS$ as follows.  If $I$ and $J$ are
ideals, generated respectively by $A$ and $B$, define
\begin{align*}
\sum(I_i\mid i\in\cI) & = \<\bigcup_{i\in\cI} A_i\>\\
I\cdot J & = \<A\odot B\> \\
1 & = \<1\>\\
0 & = \<0\>.
\end{align*}
As in Section~\ref{s:pka-star},  we need to check that these
operations are well-defined;  the analogue of Lemma~\ref{l:Twelldef}
can be seen to hold. As with $\fT$, we can extend $\fTp$ to a functor
by specifying its action on $\cat{PS}$ morphisms. If
$f:\cS\rightarrow\cS'$ is a homomorphism 
of partially additive idempotent semirings,  define $\fTp
f:\fTp\cS\rightarrow\fTp\cS'$ as $(\fTp f)(I)=\<f[I]\>$, where
$f[A]=\{f(a)\mid a\in A\}$. 
It is straightforward to check that $\fTp
f$ is a closed semiring homomorphism (see Lemma~\ref{l:fthom}). 
\fi
This gives a functor
$\fTp$ from the category $\cat{PS}$ to the category $\cat{CS}$, left
adjoint to the obvious inclusion functor
$\fPp:\cat{CS}\rightarrow\cat{PS}$. 

\begin{theorem}\label{t:PPadjoint}
The functor $\fTp$ is a left adjoint to the inclusion functor
$\fPp:\cat{CS}\rightarrow\cat{PS}$, via the adjunction map
$\phi:\Hom(\fTp\cS,\cS')\rightarrow\Hom(\cS,\fPp\cS')$ given by
$(\phi f)(x)=f(\<x\>)$.
\end{theorem}
\ifConf
\else
\begin{proof}
To show that $\fTp$ is a left adjoint to $\fPp$, we need to show that
the adjunction map $\phi$ is a natural isomorphism between
$\Hom(\fTp\cS,\cS')$ and $\Hom(\cS,\fPp\cS')$.  The proof is a
straightforward adaptation of the proof of Theorem~\ref{t:Padjoint}. 
\end{proof}
\fi

With this result, we get the adjunctions
\begin{diagram}[midshaft]
\cat{PKA}^* &  \pile{\rTo^\fC\\ \lTo_\fU} & \cat{PS} &
\pile{\rTo^\fTp\\ \lTo_\fPp} & \cat{CS}.
\end{diagram}
The composite functor $\fTp\circ\fC$ is a left adjoint to the
forgetful functor $\cat{CS}\rightarrow\cat{PKA}^*$. It is a well-known
result that any two functors that are left adjoint to the same functor
are naturally isomorphic \cite[p.85, Corollary 1]{r:maclane71}. Thus, $\fCp\circ\fT$ and $\fTp\circ\fC$ are
naturally isomorphic functors. Among other things, this implies that if
$\cK$ is a star-continuous partially additive Kleene algebra, then
$\fCp(\fT\cK)$ and $\fTp(\fC\cK)$ are isomorphic closed semirings.

\section{Conclusion}

In this paper, we introduced the notion of a partially additive Kleene
algebra, and exhibited categorical connections between the category
$\cat{PKA}^*$ of star-continuous partially additive Kleene algebras
and various other algebraic structures. 
We obtain the following commutative diagram of adjunctions:
\ifConf
\else
\begin{diagram}[midshaft]
\cat{PKA} & \pile{\rTo^\fS\\ \\ \lTo_\fK} & \cat{PKA}^* & \pile{\rTo^\fC\\ \\ \lTo_\fU} & \cat{PS} \\
  &   & \uTo^\fP \dTo_\fT  &  & \uTo^\fPp \dTo_\fTp \\
  &   & \cat{KA}^* & \pile{\rTo^\fCp\\ \\ \lTo_\fUp} & \cat{CS}.
\end{diagram}
\fi
These relationships 
provide some motivation for our definition of partially additive
Kleene algebras, by showing that partially additive Kleene algebras
fit naturally within existing algebraic structures. 

One question remains open: is there is a universal
way to complete an arbitrary partially additive Kleene algebra into a
total Kleene algebra? This asks for a left adjoint to the forgetful
functor $\cat{KA}\rightarrow\cat{PKA}$, that collapses to the
functor $\fT:\cat{PKA}^*\rightarrow\cat{KA}^*$ when restricted to
$\cat{PKA}^*$. Note that the construction in this paper only works for
star-continuous  partially additive Kleene algebras. More precisely,
the proof of Theorem~\ref{t:Padjoint} requires showing that the
adjunction $\phi$ is an isomorphism; but the well-definedness of the
inverse $\psi$ relies on the fact that $\<x\>\subseteq\{x\}^{\le}$, 
which need not hold in non-star-continuous partially additive Kleene
algebras. For instance, if $y$ is an upper bound for all $x^n$ but not
$x^*$, then $\<y\>$ will contain $x^*$ but $\{y\}^{\le}$ will not; in
other words, $\<y\>\nsubseteq\{y\}^{\le}$.

\paragraph{Acknowledgments.} Thanks to Dexter Kozen for an
enlightening comment pointing out the necessity of star-continuity.
Thanks as well to Hubie Chen for commenting on a draft of this paper.

\ifConf
\else
  \bibliographystyle{chicagor}
  \bibliography{riccardo2}
\fi

\end{document}